\documentclass[prl,aps,reprint,10pt,nofootinbib,preprintnumbers,superscriptaddress]{revtex4-2}%groupedaddress,
\pdfoutput=1
\usepackage{geometry}
\usepackage{tikz}
\usepackage{color}
\usepackage{fullpage}
\usepackage[title]{appendix}
\usepackage{hyperref}
\hypersetup{colorlinks, citecolor=Blue, linkcolor=black, urlcolor=Blue}
\usepackage{amsfonts}
\usepackage{amsmath}
\usepackage{setspace}
\usepackage[usenames, dvipsnames]{xcolor}
\usepackage{tabularx}
\usepackage{empheq}
\usepackage{comment}
\usepackage{hepunits}
\usepackage{enumitem}
\usepackage{slashed}
\usepackage{braket}
\usepackage{tensor}
\usepackage{dsfont}
\usepackage{soul}
\usepackage{booktabs}

\definecolor{light-gray}{gray}{0.9}

\newcommand{\eps}{\epsilon}

\newcommand{\be}{\begin{eqnarray}}
\newcommand{\ee}{\end{eqnarray}}
\newcommand{\nn}{\nonumber}

\newcommand{\sD}{\slashed{D}}
\newcommand{\tr} {\mathrm{tr}}
\newcommand{\Tr} {\mathrm{Tr}}
\newcommand{\dc} {\mathcal{D}}
\newcommand{\dd} {\mathrm{d}}

\newcommand{\rl}[1]{{\color{blue}[RL: #1]}}
\newcommand{\jq}[1]{\textcolor{green}{[JQ: #1]}}
\definecolor{lightviolet}{rgb}{0.75, 0.17, 0.8}

\newcommand{\am}[1]{\textcolor{purple}{[AM: #1]}}

\definecolor{darkblue}{rgb}{0.0,0.0,0.6}
\definecolor{readableAB}{rgb}{0.7,0,0.7}
\definecolor{colorRTD}{rgb}{.2,.2,.7}

\begin{document}

\bigskip\
%\preprint{}

\title{Non-Topological Chiral Anomaly: The Chiral Magnetoelectric Effect}
%Chiral Piezoelectric Effect alreay taken in literature
%Non-topological chiral anomaly
%Chiro-Electric Effect: misses the magnetic field needed

\author{R\'emy Larue}
\affiliation{School of Physical Science and Technology, ShanghaiTech University,
393 Middle Huaxia Road, Shanghai 201210, China}
\author{Amaury~Marchon}
\affiliation{Laboratoire d’Annecy de Physique Th\'eorique,
CNRS – USMB, BP 110 Annecy-le-Vieux, 74941 Annecy, France}
\author{J\'er\'emie~Quevillon}
\affiliation{Laboratoire d’Annecy de Physique Th\'eorique,
CNRS – USMB, BP 110 Annecy-le-Vieux, 74941 Annecy, France}
\author{Diego Saviot}
\affiliation{Quantum Horizons Alberta, Department of Physics and Astronomy,
University of Calgary, 2500 University Drive NW, Calgary, Alberta T2N 1N4, Canada}

\begin{abstract}
We derive the chiral anomaly in a thermal medium and show that, in addition to the well-established topological piece, the anomaly acquires a previously overlooked non-topological contribution, relevant for non-gauged axial symmetries. Remarkably, this modification of the continuity equation is independent of the electromagnetic fields, thereby giving rise to novel transport effects in relativistic hydrodynamics. We further propose an experimentally accessible new effect in Weyl semimetals which we call the Chiral Magnetoelectric Effect: an electric field proportional to the magnetic field is generated by the non-topological anomaly, even in the absence of an externally applied electric field. Beyond this concrete prediction, our results establish a broader hydrodynamic framework for non-topological chiral anomalies, opening the door to a range of previously unexplored anomaly-induced transport phenomena across condensed-matter and high-energy systems.

\end{abstract}

\maketitle

%%%%%%%% INTRODUCTION %%%%%%%%%%%

\textbf{\emph{Introduction.}}  --- 
The chiral anomaly is a well-established result, derived through various approaches and deeply connected to the topological properties of the underlying gauge theory. This often leads to considering the anomaly equation for the axial current in electric and magnetic background fields (and their axial counterparts)
\begin{equation}\label{eq:Atopo}
\langle\partial j_A\rangle=\left.{\cal A}_A\right|_{\text{topo}}=\frac{1}{2\pi^2}\left(\vec E\cdot\vec B+\frac{1}{3}\vec E_A\cdot\vec B_A\right)\;,
\end{equation}
as an immutable equation.
However, cohomologically trivial anomalies, i.e. that can be subtracted by renormalisation counterterms~\cite{Balachandran:1981cs, Andrianov:1983nd, Hu:1983ij}, can occur in addition to this topological part. When the chiral symmetry is gauged, these trivial anomalies artificially violate gauge invariance and must therefore be subtracted. However, if the axial symmetry is non-gauged, there is no motivation to subtract them.

This was first appreciated in Refs.~\cite{Balog:1984upv,Balog:1985ea} in the context of chiral perturbation theory, where an axial external flavour (non-gauged) field $A_\mu$ is introduced. The corresponding trivial chiral anomaly is~\cite{Balog:1985ea, Hu:1983ij, Andrianov:1983qj, Balachandran:1981cs}
\begin{align}
\left.\mathcal{A}_A\right|_{\text{non-topo}}&=\frac{\Lambda^2}{2\pi^2}\partial_\mu A^\mu\!+\!\frac{1}{3\pi^2}\partial_\mu(A^\mu A^2)\!-\!\frac{1}{12\pi^2}\partial^2\partial_\mu A^\mu\nn\\
&+\text{non-abelian terms}\;,\label{eq:BalogAnomaly}
\end{align}
where we displayed only the abelian part for simplicity
and $\Lambda$ is a regularization cut-off.
Note that the trivial chiral anomaly is of even intrinsic parity (i.e. even powers of Levi-Civita tensor) and is non-topological, whereas the topological part is of odd intrinsic parity.~\footnote{We use interchangeably non-topological and trivial since they are equivalent in the context of the chiral anomaly. The same applies for topological and non-trivial. This is however not true in general. For example the Weyl anomaly has a non-topological and non-trivial contribution from the Weyl tensor squared.}
The trivial anomaly is then integrated to obtain effective pions interactions, in the same way that the non-trivial chiral anomaly is integrated to obtain the Wess-Zumino-Witten term~\cite{Wess:1971yu,Witten:1983tw}. Remarkably, these effective interactions stemming from the trivial anomaly are shown to predict the decay of kaons and pions in agreement with experimental measurements~\cite{Balog:1985ea}.

Recently, it has been understood that the chiral anomaly modifies the hydrodynamics equations, leading to new transport phenomena, among which the most famous ones are the Chiral Magnetic, Vortical and Separation Effects \cite{Fukushima:2008xe,Metlitski:2005pr,Son:2009tf}.
In some of these systems, physical non-gauged axial fields appear, for instance through the lattice structure of condensed matter systems. Nevertheless, the trivial part of the chiral anomaly in hydrodynamics and condensed matter set-ups was never considered, and thus deserves to be investigated.

In the authors' previous work~\cite{Larue:2025yar, Larue:2026cjx}, a complete picture of anomaly-related transport is drawn by computing the anomaly and all the currents within the same framework: a fermion in a thermal and dense medium coupled to a vector and an axial background field is directly integrated out in the path integral. This approach does not presuppose an anomaly equation, and establishes clearly the link between anomalies and transport. For these reasons, it also provides a well-fit framework to study the trivial chiral anomaly and its associated transport.

The goal of this work is thus to systematically compute the non-topological contributions to the anomaly and currents and explore some physical implication. In the first part of the paper, the path integral formalism and Covariant Derivative Expansion in imaginary-time are introduced. In the second part, the intrinsic-parity-even currents are computed alongside the corresponding trivial anomalies. In the third part, we present a new physical effect purely due to the trivial part of the anomaly, and estimate its magnitude, concluding that it should be experimentally observable in Weyl semimetals.
Throughout this manuscript, we use the mostly-plus metric signature convention.\\

\textbf{\emph{Introduction to the formalism.}}  --- 
We consider the thermal partition function for a chiral fermion at constant temperature $T=1/\beta$,  with (internal) vector and axial chemical potentials $\mu_V^0(\vec x)$ and $\mu_A^0(\vec x)$
\begin{equation}
Z=\Tr \,e^{-\beta \int\dd^3x\,(\hat H-\mu_V^0 j^0_V-\mu_A^0 j^0_A)}\;,
\end{equation}
where $\hat H$ is the fermionic Hamiltonian density, and the vector and axial currents are
\begin{equation}
j_V^\mu=\bar\psi \gamma^\mu\psi\;,\quad j_A^\mu=\bar\psi\gamma^\mu\gamma_5\psi\;.
\end{equation}
The partition function admits a path integral representation in imaginary-time~\cite{Matsubara:1955ws,Bellac:2011kqa}
\begin{align}
&Z[\beta,\mu_V,\mu_A,\vec V,\vec A]=\int_{\text{b.c}}\dc\bar\psi\dc\psi \,e^{-\int_0^\beta \dd\tau\int\dd^3x\,{\cal L}}=\text{det}_\beta i\slashed D\;,\nn\\
&{\cal L}=\bar\psi i\slashed D\psi\;,\quad D_\mu=\partial_\mu +iV_\mu+iA_\mu\gamma_5\;.
\label{eq:Lagr}
\end{align}   
with the antiperiodic boundary conditions: $\text{b.c} = \{ \psi(i\beta,\vec{x}) = -\psi(0,\vec{x}),\;\bar\psi(i\beta,\vec{x}) = -\bar\psi(0,\vec{x}) \}$.
Conveniently, the imaginary-time formalism allows us to absorb the chemical potentials in the abelian 4-vector and 4-axial-vector fields such that $V_\mu=(\mu_V,\vec V)$ and $A_\mu=(\mu_A,\vec A)$, where $\vec V$ and $\vec A$ are the vector and axial potentials respectively; $\mu_V=\mu^0_V+\phi$ is  the electrochemical potential with $\phi$ the electrostatic potential; and likewise $\mu_A=\mu^0_A+b_0$ is the axial electrochemical potential with $b_0$ an axial electrostatic potential.

Note that the partition function $Z$ is given at a fixed reference (real) time $t_0$ and describes a local equilibrium. In~\eqref{eq:Lagr}, the path integral variables $\bar\psi$ and $\psi$ are functions of the imaginary-time only, whereas the other fields depend only on the real-time~\cite{Hongo:2016mqm,Hongo:2019rbd}. In particular, the time-derivative in $\cal L$ is an imaginary-time-derivative $\partial_0=\partial_\tau$.
As a result, the partition function $Z$ is obtained in the adiabatic regime where $\partial_{t_0}\langle \mathcal{O}\rangle\simeq\langle\partial_{t_0}\mathcal{O}\rangle\simeq 0$. Above and in the following, we omit writing the $t_0$ dependency for simplicity.

At fixed time $t_0$, the Lagrangian~\eqref{eq:Lagr} is invariant under the spatial $U(1)_V$ and $U(1)_A$ symmetries
\begin{subequations}\label{eq:U(1)VA}
\begin{alignat}{3}
&\delta^V_\theta \vec V=\vec\partial \theta(\vec x)&\;,\;&\delta^V_{\theta} \psi=e^{i\theta(\vec x)}\psi&\;,\;&\delta^V_\theta \bar\psi=\bar\psi e^{-i\theta(\vec x)}\;,\\
&\delta^A_\theta \vec A=\vec\partial \theta(\vec x)&\;,\;&\delta^A_{\theta} \psi=e^{i\theta(\vec x)\gamma_5}\psi&\;,\;&\delta^A_\theta \bar\psi=\bar\psi e^{i\theta(\vec x)\gamma_5}\;,
\end{alignat}
\end{subequations}
such that the vector and axial currents are classically conserved $\partial j_V=\partial j_A=0$.
However, the measure of the partition function may not be invariant~\cite{Fujikawa:1979ay,Fujikawa:1980eg,Fujikawa:2004cx}. If so, then the partition function is not invariant either, and we have quantum vector and axial anomalies which are given by the Jacobian of the $U(1)_{V/A}$ transformation of the path integral variables
\begin{equation}
\int_0^\beta\!\dd\tau\!\int\!\dd^3x\,\theta(\vec x)\mathcal{A}_{V/A}:=\log J_{V/A}[\theta(\vec x)]=-\delta^{V/A}_\theta \log Z\;.
\end{equation}
Since no observables depend on the imaginary time $\tau$, the $\dd\tau$ integral can be performed straightforwardly, yielding a factor $\beta$.
Using $\delta\log\det{\cal O}=\Tr\,\delta{\cal O}\,{\cal O}^{-1}$ formally, the vector and axial anomalies may be expressed as
\begin{align}\label{eq:AnTr}
\begin{split}
\!\!\!\!\!{\cal A}_{V/A}(\vec x)&=\langle \partial_i j^i_{V/A}\rangle=-\frac{1}{\beta}\frac{\delta}{\delta\theta(\vec x)}\Tr_\beta\left[(\slashed\partial\theta)\,\gamma_{V/A}\frac{1}{i\sD}\right]\;,
\end{split}
\end{align}
where we introduced the compact notation $\gamma_V=\mathds{1}$ and $\gamma_A=\gamma_5$, and the Expectation Value (EV) of an operator is defined by
\begin{equation}\label{eq:EV}
\langle{\cal O}\rangle(\vec x)=\frac{\int_{\text{b.c}}\dc\bar\psi\dc\psi \,\left(\frac{1}{\beta}\int_0^\beta\dd\tau\, {\cal O}(\tau,\vec x)\right)\,e^{-\int_0^\beta \dd\tau\int\dd^3y\,{\cal L}}}{\int_{\text{b.c}}\dc\bar\psi\dc\psi \,e^{-\int_0^\beta \dd\tau\int\dd^3y\,{\cal L}}}\;.
\end{equation}
From~\eqref{eq:AnTr} we can easily read the EV of the currents themselves
\begin{equation}
\langle j^i_{V/A}\rangle(\vec x)=\frac{1}{\beta}\frac{\delta}{\delta\theta(\vec x)}\Tr_\beta\left[\theta\,\gamma^i\gamma_{V/A}\frac{1}{i\sD}\right]\;.
\end{equation}
The anomaly and the anomalous currents can then both be similarly evaluated using the Covariant Derivative Expansion (CDE)~\cite{Gaillard:1985uh,Cheyette:1985ue,Larue:2023uyv} which was adapted to finite temperature (and curved spacetime) in~\cite{Larue:2025yar}. The introduction of the finite temperature results in a sum over Matsubara modes~\cite{Matsubara:1955ws}
\begin{equation}
\langle j^i_{V/A}\rangle=\frac{1}{\beta}\sum_{n\in\mathds{Z}}\int\frac{\dd^3q}{(2\pi)^3}\tr\,\gamma^i\gamma_{V/A}\sum_{k\in\mathds{N}}\left[\Delta_n(\vec q) (-i\sD)\right]^k\Delta_n(\vec q)\;,
\label{eq:cde_for_j}
\end{equation}
where the propagators are
\begin{equation}\label{eq:Deltan}
    \Delta_n(\vec q)=-\frac{i\gamma^0\omega_n+\gamma^i q_i}{\omega_n^2+\vec q^2}\;,
\end{equation}
with $\omega_n=\pi(2n+1)/\beta$ the Matsubara frequencies.~\footnote{The factor $i$ in~\eqref{eq:Deltan} appears because we kept the Lorentzian metric (with mostly-plus signature) instead of absorbing it in a Euclidean metric as in~\cite{Larue:2025yar}.}

The CDE used here uses the temperature as an expansion parameter, although we could as well use a mass expansion. It is straightforward to treat the case of a massive fermion, which boils down to replacing the propagator in~\eqref{eq:cde_for_j} by $\Delta_n(\vec q)=-\frac{i\gamma^0\omega_n+\gamma^i q_i+m}{\omega_n^2+\vec q^2+m^2}$. The axial anomaly expression is however modified due to the classical breaking from the mass%~\eqref{eq:explicit_breaking_by_mass}
. The vacuum result can then be recovered from the $\beta m\to\infty$ limit~\cite{Larue:2026cjx}.

The vector and axial currents are known to lead to intrinsic-parity-odd transport phenomena such as the Chiral Vortical, Separation, Magnetic Effects and others~\cite{Vilenkin_1978_vortical_current,Vilenkin:1978_blackholes,Vilenkin:1979,Fukushima:2008xe,Metlitski:2005pr,Son:2009tf,Nagaosa:2009ycg,Ilan:2019lqk} which are related to the topological chiral anomaly Eq.~\eqref{eq:Atopo} where $\vec E=\vec\partial \mu_V$, $B^i=\frac{1}{2}\eps^{ijk}\partial_j V_k$, $\vec E_A=\vec\partial\mu_A$ and $B_A^i=\frac{1}{2}\eps^{ijk}\partial_j A_k$.
These various transport effects and their relation to the chiral anomaly have been investigated in a systematic and unified manner  by the authors in~\cite{Larue:2026cjx}.
In the following, we will instead focus on the intrinsic-parity-even part of the vector and axial currents, and their contributions to the chiral anomaly as trivial anomalies.\\

\textbf{\emph{Anomalous currents and non-topological anomalies.}}  --- 
Starting from Eq (\ref{eq:cde_for_j}), we compute the vector and axial currents order by order in $k$, up to $k=3$. In the CDE approach, the sum-integrals over the Matsubara frequencies and 3-momenta yield temperature-dependent coefficients that separate from the operator part. These master-sum integrals are such that
\begin{align}
    S_{a,b,c} = \frac{1}{\beta}\sum_{n\in\mathds Z}\int\frac{\dd^d\vec q}{(2\pi)^d}\frac{(\vec q^{\,2})^{a} \omega_n^{2b}}{(\omega_n^2+\vec q^{\,2})^c} \propto T^{d+1+2(a+b-c)} \,,\nn
\end{align}
and the expansion is therefore valid for large temperatures. They are defined and computed in App.~\ref{app:MasterSumInt}.

As is usual when computing quantum anomalies, UV divergent integrals are involved. We use dimensional regularization to regularize them.
As underlined in \cite{Larue:2026cjx}, the treatment of $\gamma_5$ in dimensional regularization is crucial as it controls which symmetry, vector or axial, will be preserved. We employ the Breitenlohner–Maison–'t Hooft–Veltman (BMHV) which is known to preserve the vector gauge symmetry~\cite{Bertlmann:1996xk,Novotny:1994yx,Horejsi:1988ei,tHooft:1972tcz,Breitenlohner:1977hr,Filoche:2022dxl}, while the non-gauged axial symmetry is broken. Both will be verified explicitly.

At finite temperature, one has to compute separately the time and space components of the currents: $j_{V/A}^0$ and $j_{V/A}^i$.
At $k=1$ in the expansion~\eqref{eq:cde_for_j}, we obtain
\begin{subequations}
\label{eq:Jvk=1}
    \begin{align}
        \langle j^0_V \rangle_{|k=1} &= 4(S_{0,0,1}-2S_{0,1,2})\mu_V= -\frac{T^2}{3}\mu_V\,,\\[5pt]
        \langle j^i_V \rangle_{|k=1} &= 4\left(\left(1-\frac{2}{d}\right)S_{0,0,1}+\frac{2}{d}S_{0,1,2}\right)V^i = 0\,,
    \end{align}
\end{subequations}
with $d=3$ in the last equality. The vanishing of the spatial current is expected by gauge-invariance since the only possible contribution depends on $\vec V$ explicitly. The vector charge density $\rho_V=-j^0_V$ however picks up a non-zero expectation value, known as the Debye mass $m^2_D(T) = \frac{T^2}{3}$.~\footnote{Note that the electric/axial charge density is $\rho_{V/A}=u_\mu \langle j^\mu_{V/A}\rangle=-\langle j^0_{V/A}\rangle$ in our mostly-plus metric signature, where $u^\mu=\delta^\mu_0$ is the future-oriented timelike unit vector. } In the effective action, it corresponds to an effective mass for the vector field $\frac{1}{2}m^2_D V_0^2$~\cite{Weldon:1982aq,Bellac:2011kqa}.

For the leading order contribution to the axial current we find
\begin{subequations}\label{eq:jAk=1}
    \begin{align}
    \!\!\left.\langle j^0_A \rangle\right|_{k=1} &\!\!= 4\left(\!\left(-1+\frac{6}{d}\right) S_{0,0,1}-\frac{6}{d}S_{0,1,2}\!\right)\mu_A\!=\! -m_D(T)^2\mu_A\,,\label{eq:JA0k=1}\\[10pt]
    \!\!\left.\langle j^i_A \rangle\right|_{k=1} &\!\!= 4\left(\!\left(-1+\frac{4}{d}\right) S_{0,0,1}+\left(2-\frac{4}{d}\right)S_{0,1,2}\!\right)A^i\!=\! 0\,,
    \end{align}
\end{subequations}
Similarly to the vector current, the time component $j_A^0$ exhibits a Debye mass. Note that contrary to the vector current, an explicit $\vec A$-dependence is allowed since it is not a gauge field, although it happens to vanish here.

No contributions arise at $k=2$ in~\eqref{eq:cde_for_j} since the master sum-integrals all vanish when the numerator has an odd power of momenta or Matsubara frequencies. At $k=3$, the only non-vanishing contributions to the vector current are
\begin{subequations}
\label{eq:Jvk=3even}
\begin{align}
    \left.\langle j^0_V\rangle\right|_{\text{even}}^{k=3}=&\frac{1}{12\pi^2}\left(2\log(\beta\mu_{\text{ren}})-1\right)\vec\partial^2\mu_V\nn \\
    &-\frac{1}{\pi^2}\mu_V\mu_A^2-\frac{1}{3\pi^2}\mu_V^3 \\%\displaybreak[4]\\
    \left.\langle j^i_V\rangle\right|_{\text{even}}^{k=3}=&\frac{1}{6\pi^2}\log(\beta\mu_{\text{ren}})\partial_j F^{ij}_V\;,
\end{align}    
\end{subequations}
where we renormalized using the $\overline{\text{MS}}$-scheme (see App.\ref{app:MasterSumInt}).

In the axial current we obtain
\begin{subequations}\label{eq:jAk=3}
\begin{align}
    \left.\langle j^0_A\rangle\right|_{\text{even}}^{k=3}=&\frac{1}{6\pi^2}\log(\beta\mu_{\text{ren}})\vec\partial^2\mu_A -\frac{1}{\pi^2}\mu_A\mu_V^2-\frac{1}{3\pi^2}\vec A^2\mu_A\label{eq:JA0k=3even}\\%\displaybreak[4]\\
    \left.\langle j^i_A\rangle\right|_{\text{even}}^{k=3}=&\frac{1}{6\pi^2}\log(\beta\mu_{\text{ren}})\partial_j F^{ij}_A-\frac{1}{3\pi^2}\mu_A^2 A^i \nn\\
    &+\frac{1}{3\pi^2}\vec A^2 A^i-\frac{1}{12\pi^2}\vec\partial^2 A^i\;.
\end{align}    
\end{subequations}
Note that our result for the axial charge density differs from that of the literature~\cite{Shi:2020uyb,Yang:2020mtz}. The reason is that their currents correspond to the covariant currents as shown in App.~\ref{app:covariant_axial_charge}. The issue is that the covariant currents violate the vector gauge symmetry. Our currents, on the other hand, are the consistent currents that preserve this gauge symmetry.

The next orders $k>3$ are $\mathcal{O}(1/T)$ and thus subleading in our expansion. Although they contribute to the currents, they do not contribute to the anomaly since these terms are UV finite.

The anomalies are now obtained from the currents by taking their divergence.
Including all the contributions to the vector current Eqs.~\eqref{eq:Jvk=1},~\eqref{eq:Jvk=3even} and the intrinsic-parity-odd part~\eqref{eq:Jvk=3odd}, the divergence of the vector current vanishes such that the vector gauge symmetry is preserved~\footnote{We remind the reader that $\partial_0=\partial_\tau$ and that only the path integral variables depend on $\tau$.}
\begin{equation}
\mathcal{A}_V = 0\,.
\end{equation}
On the other hand, collecting the divergence of Eqs.~\eqref{eq:jAk=1} and~\eqref{eq:jAk=3}, the intrinsic-parity-even non-topological axial anomaly is
\begin{align}
    \begin{split}
    \left.\mathcal{A}_A\right|_{\text{non-topo}}
    &= -\frac{1}{3\pi^2} \partial_j (\mu_A^2 A^j)\\
    & +\frac{1}{3\pi^2} \partial_j (A^j A_iA^i) - \frac{1}{12\pi^2}  \partial_j\partial^j \partial_i A^i \,.
    \end{split}
    \label{eq:trivial_axial_anomaly}
\end{align}
The renormalization-dependent contributions to the vector and axial currents do not contribute to the anomaly since $\partial_i\partial_jF^{ij}_{V/A}=0$% and $\partial_0=\partial_\tau=0$
. The anomalies, even trivial, are independent of the renormalization scale as expected.

Although the anomaly can partly be recovered from the vacuum result under the substitution $A_\mu\to (\mu_A,A_i)$, the currents cannot. Indeed, the authors showed in Ref.~\cite{Larue:2026cjx} that they cannot in general be identified with the Chern-Simons currents, which also means there are more coefficients to be determined in the currents than constraints from the anomaly alone. Eq.~\eqref{eq:trivial_axial_anomaly} is in agreement with~\eqref{eq:BalogAnomaly} under $A_\mu\to (\mu_A,A_i)$ except for the cut-off dependent term $\frac{\Lambda^2}{2\pi^2}\partial_\mu A^\mu$. One could have expected that $\Lambda$ would be replaced by the cut-off of the thermal effective theory $T$, but the computation reveals that it is absent. %We also show in the massive case  that there is no term $\frac{m^2}{2\pi^2}\partial_\mu A^\mu$ either (see App.~\ref{app:massive_case}). 
In summary, the replacement $A_\mu\to(\mu_A,A_i)$ in the vacuum anomaly is a useful cross-check for certain coefficients of the anomaly, but is in no case sufficient to obtain the full anomaly and currents in a thermal and dense background.

For completeness and for the next section, we also provide the contribution from the intrinsic-parity-odd terms computed by the authors in Ref.~\cite{Larue:2026cjx}, to the vector current:
\begin{subequations}
\label{eq:Jvk=3odd}
\begin{align}
    \left.\langle j^0_V\rangle\right|_{\text{odd}}^{k=3}&=\frac{1}{2\pi^2}\vec A\cdot\vec B_V\\
    \left.\langle j^i_V\rangle\right|_{\text{odd}}^{k=3}&=\frac{1}{2\pi^2}\mu_V B^i_A-\frac{1}{2\pi^2}(\vec A\times\vec E_V)^i\;,
\end{align} 
\end{subequations}
the axial current:
\begin{subequations}
\begin{align}
    \left.\langle j^0_A\rangle\right|_{\text{odd}}^{k=3}=&\frac{1}{6\pi^2}\vec A\cdot\vec B_A\label{eq:JA0k=3odd}\\
    \left.\langle j^i_A\rangle\right|_{\text{odd}}^{k=3}=&\frac{1}{2\pi^2}\mu_V B^i_V+\frac{1}{3\pi^2}\mu_A B^i_A -\frac{1}{6\pi^2}(\vec A\times\vec E_A)^i\;,
\end{align} 
\end{subequations}
and the associated topological part of the anomaly $\left.\mathcal{A}_A\right|_{\text{topo}}$, which is written in Eq.~\eqref{eq:Atopo}.

In the following, we exhibit a new transport phenomenon due to the non-topological anomaly~\eqref{eq:trivial_axial_anomaly}, and an avenue for its experimental detection.\\

\textbf{\emph{Experimental protocol.}}  ---
Let us now propose an observable consequence of the trivial/non-topological part of the anomaly~\eqref{eq:trivial_axial_anomaly}. A well-known realization of non-gauged axial fields occurs in Weyl semimetals, where the separation between two Weyl nodes in momentum space, $\vec{b}$, and in energy, $b^0$, precisely plays that role~\cite{Ilan:2019lqk}: $A_{\mu} = (\mu_A,\vec{b})$ with $\mu_A=\mu_A^0+b^0$ as defined earlier. Weyl nodes have been reported in several materials ($\mathrm{ZrTe_5}$, $\mathrm{TaAs}$, $\mathrm{NbAs}$ for instance) and their separation has been observed with angle-resolved photoemission spectroscopy, proving that a physical realization of the axial field is feasible~\cite{Li:2014bha, Lv:2015ayh, Xu:2015cga, Xu:2015cil}. In order to have the trivial anomaly Eq.~\eqref{eq:trivial_axial_anomaly}, we need a non-constant or non-homogeneous axial field $A^{\mu}$, which can be achieved by strain or non-homogeneous magnetization~\cite{Ilan:2019lqk,Lv:2015ayh, Kamboj:2019vgc}.

\begin{figure}[h]
    \centering
    \begin{tikzpicture}[scale=3.5]
\node at (-0.1,1) []  {$V$};
\node at (-0.05,0.07) [] {0};
  \node at (1,-0.1) []  {$j$};
  \draw[-latex,thick] (0,0) -- (0,1);
  \draw[-latex,thick] (-0.7,0) -- (1,0); 
  \draw[CornflowerBlue,thick] (-0.6,0) -- (1,0.8); 
  \draw[dashed] (0.6,0.6) -- (0.8,0.6);
  \draw[dashed] (0.8,0.6) -- (0.8,0.7);
  \node at (0.9,0.64) []  {$\rho\, d$};
  \draw[dashed,<->] (0.05,0.01) -- (0.05,0.3);
  \node at (0.55,0.15) []  {$V_0 \sim \rho \,L\,\frac{\tau_V v_F^4}{T^2}(\partial^3 A)B$};
  \draw[-latex,brown,thick] (0,-0.05) -- (-0.6,-0.05);
  \node at (-0.3,-0.15) []  {$j_{\mathrm{CME}}$};
\end{tikzpicture}
\caption{Electric tension at the boundaries of the sample as a function of the electric current. The situation where $j=0$ is achieved with a non-zero potential $V_0$ that counterbalances the CME. Without the trivial anomaly, this curve crosses the origin such that no electric field persists at equilibrium.}
\label{plot:V_0}
\end{figure}
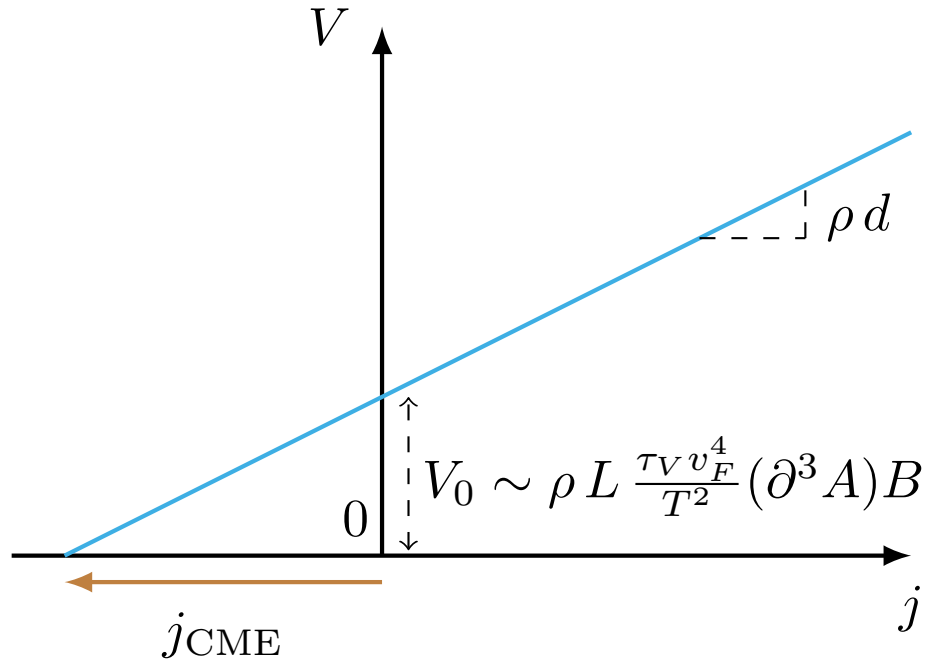

Following standard derivations of a negative magnetoresistance in Weyl semimetals induced by the chiral anomaly \cite{Li:2014bha,Grushin:2016fgr,Ong:2020ffe}, we start from the Ward identity out of equilibrium, supplemented by a phenomenological damping term involving $\tau_V$ the intervalley scattering time
\begin{equation}
\partial_t\rho_A + \partial_i \langle j^i_A\rangle = \mathcal{A}_A - \frac{\rho_A}{\tau_V}\,,
\label{eq:ward_identity_for_axial_pumping}
\end{equation}
where $\rho_A=-\langle j^0_A\rangle$ is the axial charge density.
Assuming a configuration where no axial current leaks from the boundaries, i.e. $n_i\langle j_A^i\rangle = 0$ where $\vec{n}$ is the normal to the boundary, then $\int d^3x\langle \partial_i j_A^i\rangle=0$ and the stationary solution to Eq.~\eqref{eq:ward_identity_for_axial_pumping} averaged over the sample's volume reads \footnote{Dropping the divergence of the current $\partial_i\langle j_A^i\rangle$ is justified by volume integration or by focusing on cases where it is negligible.}
\begin{equation}\label{eq:rhoA=tauA}
\rho_A = \tau_V \mathcal{A}_A\,.
\end{equation}
On the other hand, $\rho_A$ was computed in Eqs.~\eqref{eq:JA0k=1},~\eqref{eq:JA0k=3even} and~\eqref{eq:JA0k=3odd}, and reads
\begin{multline}
\rho_A = \left(\frac{T^2}{3v_F^3}+\frac{\mu_V^2}{\pi^2v_F^3} + \frac{\vec{A}^2}{3\pi^2v_F}\right)\mu_A - \frac{1}{6\pi^2}\vec{A}\cdot\vec{B}_A \\
- \frac{1}{6\pi^2v_F}\log(\beta\mu_{\mathrm{ren}})\vec{\partial}^2\mu_A\,,
\label{eq:equation_of_state_for_rho_A}
\end{multline}
where we introduced the Fermi velocity $v_F$ by restoring the dependence in $c$ then replaced by $v_F$.
We can similarly introduce the Fermi velocity in the equation~\eqref{eq:rhoA=tauA}, and including the expression of the anomaly~\eqref{eq:trivial_axial_anomaly} it reads
\begin{multline}
\frac{\rho_A}{\tau_V} = \frac{v_F}{3\pi^2}\vec{\partial}\cdot(\vec{A}\,\vec{A}^2) - \frac{v_F}{12\pi^2}\vec{\partial}^2\vec{\partial}\cdot\vec{A} - \frac{1}{3\pi^2v_F}\vec{\partial}\cdot(\mu_A^2\vec{A}) \\
 +\left.\mathcal{A}_{A}\right|_{\text{topo}}\,.
\label{eq:equation_of_anomaly_for_rho_A}
\end{multline}
where the topological part $\left.\mathcal{A}_{A}\right|_{\text{topo}}$ given by Eq.~\eqref{eq:Atopo} takes no $v_F$ factor. Eqs.~\eqref{eq:equation_of_state_for_rho_A} and~\eqref{eq:equation_of_anomaly_for_rho_A} combine together to form a differential equation that $\mu_A$ must obey. For the simple case of a constant axial chemical potential $\vec\partial\mu_A=\vec E_A=0$, it simplifies to a second order polynomial equation on $\mu_A$ which must be solved within the domain of validity of the CDE
\begin{equation}\label{eq:mu_A<<T}
T \gg |\mu_V|, |\mu_A|, ||\vec{A}||\,.
\end{equation}
For $T$ of the order of a few tens of Kelvins, i.e. $T\sim 10^{-3}\,\mathrm{eV}$, we choose $\mu_V,||\vec{A}||\sim10^{-4}\,\mathrm{eV}$. We also take typical values from Refs.~\cite{Kharzeev:2009pj,Lv:2015ayh} for the other parameters: $\tau_V\sim 10^1\,\mathrm{eV}^{-1}$, $\partial\sim 1/d$ with $d\sim 1\,\mathrm{eV}^{-1}$ the size of the sample, and $v_F\sim c/300$. Under the constraints~\eqref{eq:mu_A<<T}, we find a single solution
\begin{equation}\label{eq:muAntdominant}
\left.\mu_A \right|_{\text{non-topo}}\simeq -\frac{\tau_V v_F^4}{4\pi^2 T^2}\, \vec\partial^2\vec\partial\cdot\vec A \sim 10^{-5}\,\mathrm{eV}\,,
\end{equation}
where we only kept the dominant contribution stemming from the non-topological anomaly. Note that $\mu_A$ also takes contribution from the topological anomaly which we account for below in the conductivity~\cite{Kharzeev:2009pj,Son:2012bg, Li:2014bha, Grushin:2016fgr}. More details are deferred to App.~\ref{app:mu_Asolution}. 
This axial chemical potential creates an electric current out-of-equilibrium via the chiral magnetic effect (CME)
\begin{equation}
\vec{j}_{\mathrm{CME}} = \frac{e^2}{2\pi^2}\mu_A\vec{B}\,,
\end{equation}
which will itself trigger an electric response of the system via its Ohmic/electric conductivity $\sigma$
\begin{equation}
\vec j_V=\vec j_{\text{Ohm}}+\vec j_{\text{CME}}=\sigma\vec E+\frac{e^2}{2\pi^2}\left.\mu_A\right|_{\text{non-topo}}\vec B\;,
\end{equation}
where $\sigma=\sigma_0+\frac{3e^2\tau_V v_F^3}{4\pi^4T^2} B^2$ comes from the topological part of the anomaly~\cite{Kharzeev:2009pj, Son:2012bg, Li:2014bha, Grushin:2016fgr}. On the other hand, the net effect of the non-topological anomaly is an offset in the electric current (see Fig.~\ref{plot:V_0}).

Eventually, the system will relax to equilibrium $\vec j_V=0$, leaving a persistent electric field and electric voltage generated by the non-topological chiral anomaly
\begin{subequations}
\begin{align}
&\vec E_0=-\frac{e^2}{2\pi^2}\rho \left.\mu_A\right|_{\text{non-topo}} \vec B\sim 10^{-2} \,\text{V}/\text{m}\\
&V_0 = E_0L = -\frac{e^2}{2\pi^2}\rho\, L\, \left.\mu_A\right|_{\text{non-topo}} B \sim 10\,\mathrm{nV}\,,
\end{align}
\end{subequations}
with $ \left.\mu_A\right|_{\text{non-topo}}$ given in Eq.~\eqref{eq:muAntdominant}, and where $\rho=1/\sigma\sim 10^{-5}\,\mathrm{\Omega}. \text{m}$ and $B\sim 1\,\mathrm{T}$ are typical values found in experiments~\cite{Kharzeev:2009pj,Lv:2015ayh} and $L$ is the length of the sample in the direction of the magnetic field. The signal is amplified by taking the magnetic field in the direction of the longest size of the sample, for example $L\sim 10^2\, \text{eV}^{-1}\sim 1\,\mathrm{mm}$~\cite{Li:2014bha,Lv:2015ayh}, yielding a voltage which should be measurable in precision experiments. Note that $\vec E_0$ may be parallel or antiparallel to $\vec B$ depending on the sign of $\vec \partial^2\vec\partial\cdot\vec A$.

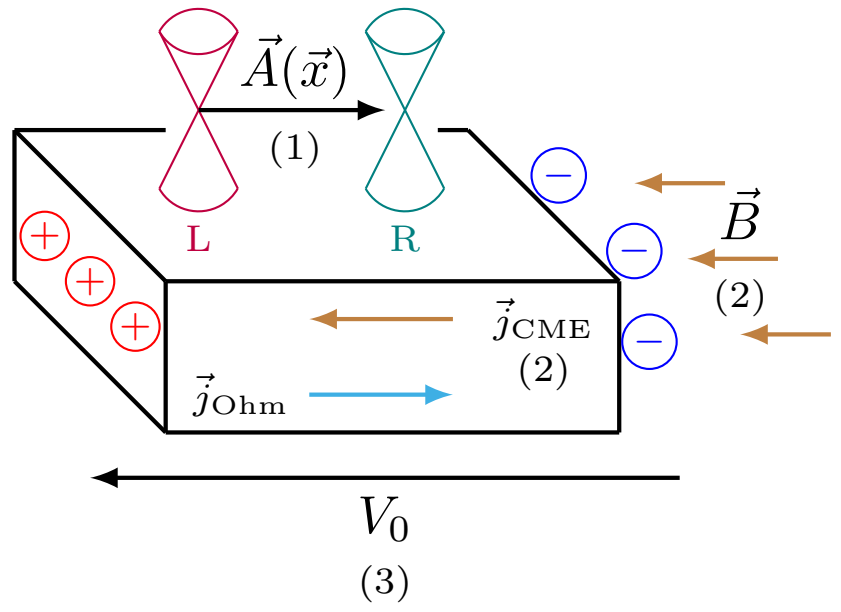
\begin{figure}[h]
\centering
\begin{tikzpicture}
% --- box geometry (cavalier projection) ---
\draw[thick] (0,0) -- (3,0);
\draw[thick] (0,1) -- (3,1);
\draw[thick] (0,0) -- (0,1);
\draw[thick] (3,0) -- (3,1);

%\draw[thick] (-1,2) -- (2,2);
\draw[thick] (-1,2) -- (0,2);
\draw[thick] (1.8,2) -- (2,2);

\draw[thick] (-1,1) -- (-1,2);
\draw[thick] (-1,1) -- (-0,0);
\draw[thick] (-1,2) -- (-0,1);
\draw[thick] (2,2) -- (3,1);

% --- Weyl cones ---
\begin{scope}[xshift=21,yshift=-54,scale=1.3]

\draw[-,purple] (-0.6,3.5) -- (-0.2,2.7);
\draw[-,purple] (-0.6,2.7) -- (-0.2,3.5);
\draw[-,purple] (-0.6,3.5) arc[start angle=-150,end angle=-30,radius=0.23];
\draw[-,purple] (-0.6,3.5) arc[start angle=150,end angle=30,radius=0.23];
\draw[-,purple] (-0.6,2.7) arc[start angle=-150,end angle=-30,radius=0.23];
\node at (-0.4,2.45) {\color{purple}\scriptsize L};

\draw[-,teal] (0.45,3.5) -- (0.85,2.7);
\draw[-,teal] (0.45,2.7) -- (0.85,3.5);
\draw[-,teal] (0.45,3.5) arc[start angle=-150,end angle=-30,radius=0.23];
\draw[-,teal] (0.45,3.5) arc[start angle=150,end angle=30,radius=0.23];
\draw[-,teal] (0.45,2.7) arc[start angle=-150,end angle=-30,radius=0.23];
\node at (0.65,2.45) {\color{teal}\scriptsize R};

\draw[-latex,thick] (-0.4,3.1) -- (0.55,3.1);
\node at (0.09,3.32) {$\vec{A}(\vec{x})$};
\node at (0.09,2.9) {\scriptsize{(1)}};
\end{scope}

% B field
\begin{scope}[xshift=0,yshift=-10,scale=1]
\draw[-latex,thick,brown] (4.4,1) -- (3.8,1);
\draw[-latex,thick,brown] (4.4-0.35,1.5) -- (3.8-0.35,1.5);
\draw[-latex,thick,brown] (4.4-0.7,2) -- (3.8-0.7,2);
\node at (4.15-0.35,1.8) {$\vec{B}$};
\node at (4.15-0.35,0.74+0.5) {\scriptsize{(2)}};
\end{scope}

% electric charges
\node[draw,circle,inner sep=0pt,minimum size=3.2mm,red] at (-0.8,1.30) {\scriptsize$+$};
\node[draw,circle,inner sep=0pt,minimum size=3.2mm,red] at (-0.5,1) {\scriptsize$+$};
\node[draw,circle,inner sep=0pt,minimum size=3.2mm,red] at (-0.2,0.7) {\scriptsize$+$};

\node[draw,circle,inner sep=0pt,minimum size=3.6mm,blue] at (3.2,0.6) {\scriptsize$-$};
\node[draw,circle,inner sep=0pt,minimum size=3.6mm,blue] at (3.1,1.2) {\scriptsize$-$};
\node[draw,circle,inner sep=0pt,minimum size=3.6mm,blue] at (2.6,1.7) {\scriptsize$-$};

% electric current j_CME and j_Ohm
\draw[-latex,thick,brown] (1.9,0.75) -- (0.95,0.75);
\node at (2.5,0.75) {\scriptsize $\vec{j}_{\mathrm{CME}}$};
\node at (2.5,0.4) {\scriptsize{(2)}};

\draw[-latex,thick,CornflowerBlue] (0.95,0.25) -- (1.9,0.25);
\node at (0.5,0.25) {\scriptsize $\vec{j}_{\mathrm{Ohm}}$};

% V_0 potential
\draw[-latex,thick] (3.4,-0.3) -- (-0.5,-0.3);
\node[below] at (1.45,-0.3) {$V_0$};
\node at (1.45,-1) {\scriptsize{(3)}};
\end{tikzpicture}
\caption{(1): A non-homogeneous axial field $\vec{A}(\vec{x})$ sources the trivial chiral anomaly, which creates an axial chemical potential $\mu_A$. (2): Placed in a magnetic field $\vec{B}$, the axial chemical potential fuels the chiral magnetic effect $\vec{j}_{\mathrm{CME}}$. (3): The electric current $\vec{j}_{\mathrm{CME}}$ generates an accumulation of charge at the boundaries and an electric tension $V_0$ along with a corresponding Ohmic current $\vec{j}_{\mathrm{Ohm}}$ counterbalancing the CME.}\label{fig:setup}
\end{figure}

We emphasize that this electric potential appears even in the absence of externally applied electric field. It is purely generated by the non-topological chiral anomaly in the presence of an external magnetic field and in Weyl semimetals whose Weyl nodes are separated in momentum space (see Fig.~\ref{fig:setup}). We thus call it the Chiral Magnetoelectric Effect (CMEE).

The evaluation of the non-topological anomaly and of the CMEE could use further refinements.
In particular, we believe that an out-of-equilibrium study would be highly beneficial to understanding the details of the effects. This would also enable one to include the effect of a time-dependent axial field, which can be generated in Weyl semimetals~\cite{Cortijo:2015hlt,Pikulin:2016wfj,Hannukainen:2019ipg,Araki:2019jfv,SCHMELTZER2023169380}, and is also expected to contribute to the non-topological anomaly. Moreover, the role of the physical boundaries of the system may be impactful, as shown in Refs.~\cite{Gorbar:2015wya,Valgushev:2015pjn,Pikulin:2016wfj,Sitenko:2016iqb,Okuma:2016,Gorbar:2018vuh,Buzzegoli:2023eeo} for the chiral magnetic effect. It would be interesting to quantify their effect on the CMEE for a better description of the mechanism of axial charge accumulation.\\ 

\textbf{\emph{Discussion and conclusions.}} ---
In this paper, we computed the vector and axial currents and their associated anomaly in a thermal and dense medium. For the first time, we accounted for the non-topological (i.e. cohomologically trivial) part of the anomaly, which is relevant to systems with non-gauged axial fields. The non-topological anomaly enters the divergence of the axial current in the presence of a (non-gauged) non-constant axial field alongside the usual topological anomaly. 
In particular, let us point out that in the bottom-up approaches found in the literature~\cite{Son:2009tf,Neiman:2010zi, Yang:2022ksq, Hidaka:2017auj}, the continuity equation~\eqref{eq:Atopo} including only the topological anomaly is used as a matching condition. Our results show, however, that this equation is incomplete, and that the matching should be done onto the total anomaly including the non-topological part Eq.~\eqref{eq:trivial_axial_anomaly} in the presence of a non-gauged axial field. Such a scenario arises for example in Weyl semimetals, thus calling for a re-evaluation of the physical implications of the chiral anomaly. We emphasize that the non-topological anomaly does not simply lead to new terms in the vector and axial currents, but directly modifies the continuity equation. The importance of this finding in the active field of chiral transport and Weyl semimetals deserves further investigation, with potential new effects to be discovered.

We show an example of such a new effect purely generated by the non-topological part of the anomaly, that we call the Chiral Magnetoelectric Effect (CMEE). It is the appearance of a persistent electric field at equilibrium in the presence of a non-constant axial field and external magnetic field, even in the absence of an externally applied electric field. Our order of magnitude analysis, based on experimentally achievable values of the parameters, reveals that the CMEE should be measurable in precision experiments.\\

\textit{Acknowledgements.} --- We thank M. Chernodub, C. Delaunay, A. Grushin, K. Landsteiner, E. Mottola and P.N.H. Vuong for helpful discussions. The work of A.M. and J.Q. is supported by the EFFORT project, funded through the IRGA program of Université Grenoble Alpes (UGA), and by the Tremplin project from CNRS Physique. The work of R.L is supported by the Science and Technology Commission of Shanghai Municipality (grant No. 24ZR1450600). The work of D.S. is supported by the Quantum Horizons Alberta initiative and the University of Calgary.

\bibliographystyle{utphys}
\bibliography{biblio}

\clearpage
\appendix

\setcounter{secnumdepth}{2}
\onecolumngrid

In this Supplementary Material, we present the computation of the currents and anomalies for massive fermions, in the vacuum and at finite temperature. We also detail the computation of the master sum-integrals, the estimation of $\mu_A$ for the CMEE, and the computation of the covariant axial charge conservation.

\section{Master sum-integrals}
\label{app:MasterSumInt}

The master sum-integrals are defined as:
\begin{align}
    S_{a,b,c} = \frac{1}{\beta}\sum_{n\in\mathds Z}\int\frac{\dd^d\vec q}{(2\pi)^d}\frac{\vec q^{2a} \omega_n^{2b}}{(\omega_n^2+\vec q^2 + m^2)^c} \,,
\end{align}
with $d=3-\epsilon$ in dimensional regularization.

The index $a$ is first lowered to zero using
\begin{align}
    S_{a,b,c} = S_{a-1,b,c-1}-S_{a-1,b+1,c}-m^2 S_{a-1,b,c} \;,\quad\text{for  }\, a>0\;,
\end{align}
master sum-integrals are then expressed in terms of $S_{0,b,c}$, which is evaluated using a Mellin transformation~\cite{Larue:2026cjx}
\begin{align}
\begin{split}
    S_{0,b,c}&=\frac{2^{1-d}}{\pi\Gamma (c)}m^{1-s_0}\sum_{k\in\mathds N} \frac{(-1)^k}{k!} \left(1-2^{-s_k}\right) \pi ^{-\frac{d}{2}-s_k} \Gamma \left(c-\frac{d}{2}+k\right) \zeta (s_k) (\beta  m)^{s_k-1}\,,
    \end{split}
\end{align}
where $s_k = 2(c-b) - d + 2k$. In the massless case, the power series reduces to one term, and we give those relevant to our computation in Tab.~\ref{tab:sum-integrals}.

UV divergences  appear in the limit $\zeta(1+\eps)\rightarrow\frac{1}{\eps}+\gamma_E$ with $\gamma_E$ the Euler constant. After accounting for the finite terms of the form $\eps/\eps$, the remaining divergences are renormalized in the $\overline{\text{MS}}$-scheme, i.e.
\begin{equation}
c'_{div} = \frac{1}{2\pi^2}\left(\frac{2}{\eps}+\gamma_E+\log(4/\pi)+\log\left(\beta^2\mu^2_{\rm ren}\right)\right)\, ~~~\text{is replaced by } ~~~ \frac{1}{2\pi^2} \log\left(\beta^2\mu^2_{\mathrm{ren}}\right)\,.
\end{equation}

\begin{table}[h!]
\centering
\renewcommand{\arraystretch}{2.2}   
\setlength{\tabcolsep}{10pt}        
\begin{tabular}{|c|c|c|c|}
\hline
$c \backslash b$ & 0 & 1 & 2 \\ \hline
1 & $-\frac{T^2}{24}$ & $\frac{7 \pi^2 T^4}{240}$ & $-\frac{31 \pi^4 T^6}{504}$ \\ \hline
2 & $\frac{1}{8 \pi^2 \epsilon} +\frac{r}{16 \pi^2}$ & $\frac{T^2}{48 }$ & $-\frac{7 \pi^2T^4}{480}$ \\ \hline
3 & $\frac{7\zeta(3)}{128 \pi^4T^2}$ & $\frac{1}{32 \pi^2 \epsilon} +\frac{r+2}{64 \pi^2}$ & $\frac{T^2}{192}$ \\ \hline
4 & $\frac{31 \zeta(5)}{1024 \pi^6 T^4}$ & $\frac{7 \zeta(3)}{256 \pi^4 T^2}$ & $\frac{1}{64 \pi^2 \epsilon} +\frac{3r + 8}{384 \pi^2}$ \\ \hline
\end{tabular}
\caption{$S_{0,b,c}$ for $m=0$, with $d\rightarrow3-\epsilon$, $r=\gamma_E+\log 4\pi +\log(\beta^2\mu_{\mathrm{ren}}^2)$, $\mu_{\mathrm{ren}}$ the renormalisation scale and $\gamma_E$ the Euler constant.}
\label{tab:sum-integrals}
\end{table}

\section{Solutions to $\mu_A$}
\label{app:mu_Asolution}

From Eqs.~\eqref{eq:equation_of_state_for_rho_A} and~\eqref{eq:equation_of_anomaly_for_rho_A}, and for $\partial \mu_A=0$, we obtain a degree-two polynomial equation on $\mu_A$:
\begin{equation}
a\,\mu_A^2 + b\,\mu_A + c = 0\,,
\label{eq:poly_for_muA}
\end{equation}
with
\begin{align}
a &= -\frac{\tau_V}{3\pi^2v_F}\vec{\partial}\cdot\vec{A} \,{\sim 10^{-1}\,\mathrm{eV}}\;, \\
b &= -\underset{\sim 1\,\mathrm{eV}^2}{\frac{T^2}{3v_F^3}} - \underset{\sim 10^{-2}\,\mathrm{eV}^2}{\frac{\mu_V^2}{\pi^2v_F^3}} - \underset{\sim 10^{-5}\,\mathrm{eV}^2}{\frac{1}{3\pi^2v_F}\vec{A}^2}\;, \\
c &= \underset{\sim 10^{-13}\,\mathrm{eV}^3}{\frac{\tau_V v_F}{3\pi^2}\vec\partial\cdot(\vec A \,\vec A^2)} - \underset{\sim 10^{-5}\,\mathrm{eV}^3}{\frac{\tau_V v_F}{12\pi^2}\vec\partial^2 \vec\partial\cdot\vec A} +\underset{\sim 10^{-8}\,\mathrm{eV}^3}{\frac{1}{6\pi^2}\vec{A}\cdot\vec{B}_A} + \tau_V\left.\mathcal{A}_A\right|_{\text{topo}}\;,
\end{align}
where $\left.\mathcal{A}_A\right|_{\text{topo}}=\frac{1}{2\pi^2}\left(\vec E\cdot \vec B+\frac{1}{3}\vec E_A\cdot\vec B_A\right)$ with $\vec E_A=\vec\partial\mu_A=0$. The numerical estimations correspond to the absolute value of each quantity with the value of the parameters chosen in the experimental protocol described in the manuscript. Note that $\vec E$ is not an externally applied electric field, it is potentially generated in the system by the distribution of charges at equilibrium and hence included at that stage.

For the solution $\mu_A$ to be within the regime of validity of the expansion, we must have $|\mu_A|\ll T\sim 10^{-3}$eV. We thus look for a solution such that $|\mu_A|\lesssim 10^{-4}$eV, and~\eqref{eq:poly_for_muA} becomes
\begin{equation}
0 = \underset{\lesssim 10^{-9}\,\mathrm{eV}^3}{a\mu_A^2} + \underset{\lesssim 10^{-4}\,\mathrm{eV}^3}{b\mu_A} + \underset{\sim 10^{-5}\,\mathrm{eV}^3}{c} \simeq b\mu_A + c\,.
\end{equation}
We thus find the approximate solution
\begin{equation}
\mu_A \simeq -\frac{c}{b} \simeq -\frac{\tau_V v_F^4}{4\pi^2T^2}\vec{\partial^2}\vec{\partial}\cdot\vec{A} + \frac{3\tau_Vv_F^3}{T^2}\mathcal{A}_{A}|_{\mathrm{topo}}\,.
\end{equation}
Finally, at equilibrium, we find a persistent electric field generated by the trivial anomaly
\begin{align}
\vec{j}_V =0 &= \vec{j}_{\mathrm{Ohm}} + \vec{j}_{\text{CME}} \\
&= \sigma_0\vec{E}_0 + \frac{e^2}{2\pi^2}\mu_A\vec{B} \\
&= \left(\sigma_0 + \frac{3 e^2\tau_Vv_F^3}{4\pi^4T^2}\vec{B}\,\vec{B}\cdot\right)\vec{E}_0 - \frac{e^2\tau_V v_F^4}{8\pi^4T^2}(\vec{\partial^2}\vec{\partial}\cdot\vec{A})\vec B\,,
\end{align}
which is solved by
\begin{equation}
\vec E_0=\frac{e^2\tau_V v_F^4 \rho}{8\pi^4T^2}\,(\vec{\partial^2}\vec{\partial}\cdot\vec{A})\vec B\;.
\end{equation}
where $\sigma=\sigma_0+\frac{3e^2\tau_Vv_F^3}{4\pi^4T^2}B^2$ and $\rho=1/\sigma$, and consequently
\begin{equation}
|\mu_A(E_0)|\sim 10^{-5}\text{eV}\ll T\;.
\end{equation}

\section{Covariant axial charge density}\label{app:covariant_axial_charge}

Instead of working with the partition function $Z=\det_\beta i\slashed D$, we use its modulus
\begin{equation}
|Z|=|\text{det}_\beta i\slashed D|=\sqrt{\text{det}_\beta ((i\slashed D)^\dagger)\text{det}_\beta (i\slashed D)}=\sqrt{\text{det}_\beta((i\slashed D)^\dagger i\slashed D)}\;.
\end{equation}
The currents and anomalies derived from the modulus of the partition function are known to yield the covariant form of the currents and anomalies~\cite{Fujikawa:1983bg,Alvarez-Gaume:1983ihn,Banerjee:1986bu,Ball:1988xg,Bertlmann:1996xk,Larue:2023uyv}. Indeed, the modulus of $i\slashed D$ naturally lets the curvatures appear
\begin{equation}
(i\slashed D)^\dagger i\slashed D=(i\slashed\partial-\slashed V+\slashed A\gamma_5)(i\slashed\partial-\slashed V-\slashed A\gamma_5)=D^2+\sigma\cdot F\;,
\end{equation}
where $\sigma^{\mu\nu}=\frac{1}{4}[\gamma^\mu,\gamma^\nu]$, and $F_{\mu\nu}=iF^V_{\mu\nu}+iF^A_{\mu\nu}\gamma_5$, $F^{V}_{\mu\nu}=\partial_\mu V_\nu-\partial_\nu V_\mu$ and $F^{A}_{\mu\nu}=\partial_\mu A_\nu-\partial_\nu A_\mu$.
The effective action $W=-\log Z$ is then expanded using the CDE at finite temperature~\cite{Larue:2026cjx} in the bosonic form
\begin{equation}
W =\frac{1}{2}\int_0^\beta \dd\tau\int\dd^3x\frac{1}{\beta}\sum_{n\in\mathds{Z}}\int\frac{\dd^3q}{(2\pi)^3}\sum_{k\geq1}\frac{1}{k}\tr\Big[\Delta_n\left(- D^2-2i Q_n\cdot D-\sigma\cdot F\right)\Big]^k\;,
\end{equation}
where $Q_{n,\mu}=(i\omega_n,\vec q)$, $\Delta_n(\vec q)=-1/(\omega_n^2+\vec q^2)$.
Focusing on the case where $\vec{V}=\vec{A} = \vec{\partial}\mu_A = \vec{\partial}\mu_V = 0$ for simplicity, we find the leading order terms:
\begin{equation}
W=-\frac{T^2}{6}(\mu_V^2+\mu_A^2)-\frac{1}{12\pi^2}(\mu_A^4+6\mu_V^2\mu_A^2+\mu_V^4)+\mathcal{O}(\beta)\;,
\end{equation}
where the master sum-integrals defined and computed in App.~\ref{app:MasterSumInt} were used.
Finally we derive with respect to the  axial chemical potential to obtain the covariant axial charge density
\begin{equation}
\rho_A=-\langle j^0_A\rangle=-\frac{\delta W}{\delta\mu_A}=\frac{T^2}{3}\mu_A+\frac{1}{3\pi^2}\mu_A^3+\frac{1}{\pi^2}\mu_V^2\mu_A+\mathcal{O}(\beta)\;,
\end{equation}
which is in agreement with Refs.~\cite{Shi:2020uyb,Yang:2020mtz} and confirms that they used the covariant currents which violate the vector gauge symmetry.

\end{document}